\documentclass[showpacs,floats,floatfix,superscriptaddress,aps,pra]{revtex4}

\usepackage{eurosym}
\usepackage{amsfonts}
\usepackage{amssymb}
\usepackage{amsmath}
\usepackage{graphicx}
\usepackage{bm}
\usepackage{color}
\usepackage{epsfig}
\usepackage{calc}
\usepackage{ifthen}
\usepackage{blindtext}

\makeatletter
\newcommand{\thickhline}{%
    \noalign {\ifnum 0=`}\fi \hrule height 1pt
    \futurelet \reserved@a \@xhline
}
\makeatother

\begin{document}


\title{Achromatic polarization rotator with tunable rotation angle}

\date{\today }

\email{e.dimova@issp.bas.bg} 

\begin{abstract}
We theoretically suggest and experimentally demonstrate a broadband composite optical rotator that is capable of rotating the polarization plane of a linearly-polarized light at any chosen angle. The  device is composed of an even number of half-wave plates rotated at specific angles with respect to their fast-polarization axes. The frequency bandwidth of the polarization rotator in principal increases with the number of half-wave plates. Here we experimentally examine the performance of rotators composed of two, four, six, eight and ten half-wave plates.
\end{abstract}

\author{Elena Stoyanova}
\affiliation{Department of Physics, Sofia University, James Bourchier 5 Blvd, 1164 Sofia, Bulgaria}
\author{Mouhamad Al-Mahmoud}
\affiliation{Department of Physics, Sofia University, James Bourchier 5 Blvd, 1164 Sofia, Bulgaria}
\author{Hristina Hristova}
\affiliation{Institute of Solid State Physics, Bulgarian Academy of Sciences, 72 Tsarigradsko chauss\'{e}e Blvd., 1784 Sofia, Bulgaria}
\author{Andon Rangelov}
\affiliation{Department of Physics, Sofia University, James Bourchier 5 Blvd, 1164 Sofia, Bulgaria}
\author{Emilia Dimova}
\affiliation{Institute of Solid State Physics, Bulgarian Academy of Sciences, 72 Tsarigradsko chauss\'{e}e Blvd., 1784 Sofia, Bulgaria}

\author{Nikolay V. Vitanov}
\affiliation{Department of Physics, Sofia University, James Bourchier 5 Blvd, 1164 Sofia, Bulgaria}

\maketitle


\section{Introduction}


Broadband polarization manipulation of light has been a topic of great interest in optics for many years \cite{West, Destriau, Pancharatnam1, Pancharatnam2, Harris1, Harris2, Peters, Rangelov, Dimova, Messaadi}.
First achromatic wave plates were developed in combinations of plates having different birefringence dispersions \cite{West}, then achromatic
retarders composed of two and three wave plates of the same material but different thicknesses were demonstrated \cite{Destriau, Pancharatnam1, Pancharatnam2}.
Later, achromatic wave plates with six \cite{Harris1}, ten \cite{Harris2} and arbitrary number wave plates \cite{Peters} were experimentally implemented.
In contrast to achromatic wave plates, which have long history, the achromatic polarization rotators were proposed \cite{Rangelov} and experimentally realized \cite{Dimova, Messaadi} only recently.
The demonstrated broadband polarization rotator schemes heretofore use two achromatic half-wave plates with the rotation angle being double the angle
between the fast optical axis of the two half-wave plates \cite{Rangelov, Dimova, Messaadi}.
This approach uses the universal principle that two crossed half-wave plates serve as a polarization rotator, and therefore any combination of two achromatic half-wave plates serve as a broadband polarization rotator.
In the case of the Messaadi et al. \cite{Messaadi} the achromatic half-wave plates were implemented with two double Fresnel rhombs, while in the case of \cite{Rangelov,Dimova} the achromatic half-wave plates were composite half-wave plates \cite{Peters,Ivanov}.

In this paper, we further develop the idea of broadband polarization rotator by using a set of an even number of half-wave plates rotated at predetermined angles.
The even number half-wave plates is crucial for the proposed rotator due to the fact that combination of any two rotators is a rotator, therefore we arrange a sequences of rotators, each of which is combination of two half-wave plates.
The realized rotators are theoretically predicted by the use of the additional free parameters in order to perform of the rotator bandwidth.


\begin{figure}
\includegraphics[width=1\columnwidth]{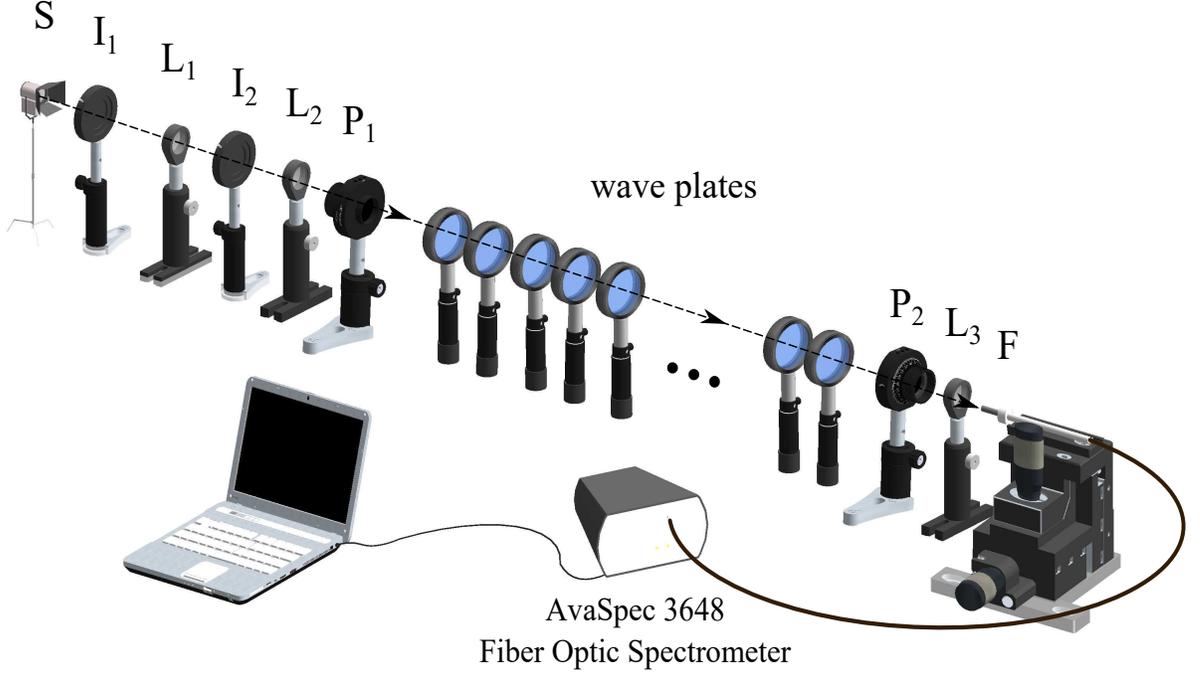}
\caption{Experimental setup. A collimated beam of white light is formed by the light source $S$, irises $I_{1}$ and $I_{2}$, and lenses $L_{1}$ and $L_{2}$. After the polarizer $P_{1}$ the light is linearly polarized in the vertical direction and passes through the broadband optical rotator, built by WPs. The second polariser $P_{2}$ is used as an analyser. With the help of lens $L_{3}$ the beam is focused onto the optical fibre entrance $F$ which is connected to a spectrometer and a computer.}
\label{fig1}
\end{figure}


\section{Theory}


The wave plates (or retarders) and rotators are two of the basic elements in polarization optics \cite{Hecht,Yariv}.
In the horizontal-vertical (HV) basis a rotation at an angle $\theta $ is described by the Jones matrix
\begin{equation}
R\left( \theta \right) =\left[\begin{array}{cc}
\cos \theta  & \sin \theta  \\
-\sin \theta  & \cos \theta
\end{array}\right] .
\end{equation}%
The combination of two rotators is also a rotator,
\begin{equation}
R\left( \theta _{2}\right) R\left( \theta _{1}\right) =R\left( \theta_{1}\right) R\left( \theta _{2}\right) =R\left( \theta _{1}+\theta_{2}\right) .  \label{combination of two rotators}
\end{equation}
In the HV basis, the Jones matrix for a retarder when the optical axes are aligned with the horizontal and vertical directions is given by
\begin{equation}
J\left( \varphi \right) =\left[\begin{array}{cc}
e^{i\varphi /2} & 0 \\ 0 & e^{-i\varphi /2}
\end{array}\right] ,
\end{equation}
where $\varphi =2\pi L(n_{\mathnormal{f}}-n_{\mathnormal{s}})/\lambda $ is the phase shift, with $\lambda $ representing the vacuum wavelength, $n_{\mathnormal{f}}$ and $n_{\mathnormal{s}}$ the refractive indices along the fast and slow axes, respectively, and $L$ is the thickness of the retarder plate.
The most commonly used retarders are the half-wave plates ($\varphi =\pi $) and the quarter-wave plates ($\varphi =\pi /2$).
If the fast and slow optical axes of the retarder are rotated at an angle $\theta $ with respect to the coordinates of the HV basis then the Jones matrix $J$ is given as
\begin{equation}
J_{\theta }\left( \varphi \right) =R\left( -\theta \right) J\left( \varphi\right) R\left( \theta \right) ,  \label{jones2}
\end{equation}
or explicitly
\begin{subequations}
\label{retarder}
\begin{align}
J_{\theta }\left( \varphi \right) _{11} &= e^{i\varphi /2}\cos ^{2}\left(\theta \right) +e^{-i\varphi /2}\sin ^{2}\left( \theta \right) , \\
J_{\theta }\left( \varphi \right) _{12} &= -i\sin \left( 2\theta \right)\sin \left( \varphi /2\right) , \\
J_{\theta }\left( \varphi \right) _{21} &= -i\sin \left( 2\theta \right)\sin \left( \varphi /2\right) , \\
J_{\theta }\left( \varphi \right) _{22} &= e^{-i\varphi /2}\cos ^{2}\left(\theta \right) +e^{i\varphi /2}\sin ^{2}\left( \theta \right) .
\end{align}
\end{subequations}

Now let us consider a sequence of two half-wave plates rotated at angles $\theta_1$ and $\theta_2$ with respect to the HV basis.
We multiply the Jones matrices of the two half-wave plates ($\varphi =\pi $) given in Eq. (\ref{jones2}), to obtain the total propagator
\begin{equation}
J_{\theta _{1}}(\pi )J_{\theta _{2}}(\pi )=-\left[\begin{array}{cc}
\cos \left( 2\left( \theta _{2}-\theta _{1}\right) \right)  & \sin \left(2\left( \theta _{2}-\theta _{1}\right) \right)  \\
-\sin \left( 2\left( \theta _{2}-\theta _{1}\right) \right)  & \cos \left(2\left( \theta _{2}-\theta _{1}\right) \right)
\end{array}\right] ,  \label{rotator2}
\end{equation}
which is a Jones matrix for a rotator up to an unimportant minus sign.
For a sequence of $N$ such pairs the Jones matrix is
\begin{equation}
\mathbf{J}_{\alpha }(\pi ) = [J_{\theta _{1}}(\pi )J_{\theta _{2}}(\pi)] [J_{\theta _{3}}(\pi )J_{\theta _{4}}(\pi)] \cdots [J_{\theta _{2N-1}}(\pi ) J_{\theta _{2N}}(\pi )] .
\end{equation}
By using the property of Eq. (\ref{combination of two rotators}) we obtain
\begin{equation}
\mathbf{J}_{\alpha }(\pi )
=\left[\begin{array}{cc}
\cos \left( \alpha \right)  & \sin \left( \alpha \right)  \\
-\sin \left( \alpha \right)  & \cos \left( \alpha \right)
\end{array}\right] ,
\end{equation}
which is a rotator with a rotation angle $\alpha $ given as
\begin{equation}
\alpha =2\sum_{k=1}^{2N}\left( -1\right) ^{k}\theta _{k}.
\end{equation}
It is easy to see that we can have the same rotation matrix $\mathbf{J}_{\alpha }(\pi )$ if each individual rotator is rotated at an additional angle,
\begin{align}
\mathbf{J}_{\alpha }(\pi ) &= [J_{\theta _{1}+\delta _{1}}(\pi)J_{\theta _{2}+\delta _{1}}(\pi )]
[J_{\theta _{3}+\delta _{2}}(\pi)J_{\theta _{4}+\delta _{2}}(\pi )] \cdots \notag\\
 &\cdots [J_{\theta _{2N-1}+\delta _{N}}(\pi ) J_{\theta _{2N}+\delta _{N}}(\pi )].
\end{align}
Therefore we can use the additional angles $\delta _{1},\delta _{2}, \ldots \delta _{N}$ as free parameters to optimise the bandwidth performance of our rotator.

We now define the fidelity $\mathfrak{F}$,
\begin{equation}
\mathfrak{F}\left( \varepsilon \right) =\frac{1}{2}\left\vert Tr\left(\mathbf{R}^{-1}(\alpha )\mathbf{J}_{\alpha }(\pi +\varepsilon )\right)
\right\vert .
\end{equation}
If the two operators $\mathbf{R}(\alpha )$ and $\mathbf{J}_{\alpha }(\pi+\varepsilon )$ are identical then $\mathfrak{F}=1$, but if the two matrices differ from one another then fidelity drops.
Here $\varepsilon $ represents the systematic deviation from the half-wave plate.
Obviously, for the central wave length at which the wave plates serve as half-wave plates, we have $\varepsilon =0$ and $\mathfrak{F}\left( 0\right) =1$.

\begin{table}[bth]
\caption{Calculated angles of rotation $\protect\theta _{N}$ (in degrees) for different numbers N of constituent half-wave plates and different rotator angles.}
\begin{tabular}{c c} \\ \hline
& 15 Degree Rotator\\ \hline
N & Rotation angles $\theta _{1},\theta _{2},...\theta _{N}$ \\ \hline
2 & (-3.75; 3.75) \\
4 & (40.6; 119.3; 116.4; 22.7) \\
6 & (64.7; 112.9; 58.6; 43.0; 99.6; 52.1) \\
8 & (-33.9; 165.7; 175.6; 73.5; 169.5; 89.1; 65.5; 33.5) \\
10 & (75.4; 4.9; 57.4; 56.9; 6.0; 124.4; 156.6; 73.1; 35.3; 56.4) \\
\end{tabular}

\begin{tabular}{c c}
\hline
& 30 Degree Rotator\\ \hline
N & Rotation angles $\theta _{1},\theta _{2},...\theta _{N}$ \\ \hline
2 & (-7.5; 7.5) \\
4 & (183.0;176.4; 78.8; 70.4) \\
6 & (119.2; 110.8; 55.4; 103; 83.14; 29.0) \\
8 & (129.4; 166.3; 51.4; 4.7; 79.5; 128.2; 63.1; 9.2) \\
10 & (43.4; 125.7; 114.6; 172.2; 112.2; 99.4; 156.0; 55.6; 141.3; 99.4) \\
\end{tabular}

\begin{tabular}{c c}
\hline
& 45 Degree Rotator\\ \hline
N & Rotation angles $\theta _{1},\theta _{2},...\theta _{N}$ \\ \hline
2 & (-11.25; 11.25) \\
4 & (106.90; 92.68; 170.96; 162.68) \\
6 & (46.9; 173.2; 47.2; 22.4; 148.8; 24.9) \\
8 & (-117.7; 16.3; 97.2; 115.5; 173.2; 109.1; 175.4; 64.8) \\
10 & (123.4; 50.8; 80.6; 175.1; 49.6; 61.7; 172.1; 22.6; 48.7; 161.3) \\  \hline
\end{tabular}
\label{table1}
\end{table}

\begin{table}[tbh]
\caption{Calculated angles of rotation $\protect\theta _{N}$ (in degrees) for different numbers N of constituent half-wave plates and different rotator angles.}
\begin{tabular}{c c}
\hline
& 60 Degree Rotator\\ \hline
N & Rotation angles $\theta _{1},\theta _{2},...\theta _{N}$ \\ \hline
2 & (-15; 15) \\
4 & (138.2; 29.3; 9.9; 88.9) \\
6 & (14.5; 138.2; 14.3; 172.2; 113.9; 162.3) \\
8 & (-43.4; 23.5; 121.0 179.7; 122; 12.6; 56.6 10.4) \\
10 & (129.0; 88.2; 154.4; 24.9; 105.9; 131.6; 63.0; 110.5; 54.6; 121.7) \\
\end{tabular}

\begin{tabular}{c c}
\hline
& 75 Degree Rotator\\ \hline
N & Rotation angles $\theta _{1},\theta _{2},...\theta _{N}$ \\ \hline
2 & (-18.75; 18.75) \\
4 & (61.9; 133.8; 117.4; 8.1) \\
6 & (164.0; 119.1; 179.8; 132.0; 75.5; 130.7) \\
8 & (111.4; 60.0; 56.1; 133.2; 128.3; 51.7; 126.9; 140.2) \\
10 & (255.8; 170.3; 54.3; 67.3; 64.3; 98.6; 17.3; 3.9; 37.9; 51.9) \\
\end{tabular}

\begin{tabular}{c c}
\hline
& 90 Degree Rotator\\ \hline
N & Rotation angles $\theta _{1},\theta _{2},...\theta _{N}$ \\ \hline
2 & (-22.5; 22.5) \\
4 & (188.7; 77.5; 57.9; 124.2) \\
6 & (126.3; 116.0; 164.8; 90.1; 63.1; 103.0) \\
8 & (264.9; 84.7; 22.3; 87.5; 125.8; 131.3; 65.4; 129.9) \\
10 & (31.1; 38.9; 142.9; 3.4; 72.0; 2.7; 6.9; 116.2; 146.4; 193.2; 31.1) \\  \hline
\end{tabular}
\label{table2}
\end{table}

In order to find the optimized angles of rotation of each wave plate we use the Monte Carlo method and for each number of wave plates and each rotator angles and we generate 10$^{4}$ sets of random angles $\theta _{1},\theta_{2},\ldots ,\theta _{2N}$.
We pick up solutions, which in the interval of $\varepsilon \in \left[ -\pi ,\pi \right] $ deliver the biggest area of the fidelity $\mathfrak{F}\left( \varepsilon \right) $ and also ensure a flat top.
The angles are presented in Tables \ref{table1} and \ref{table2}.

We note that by using the numerical solution of many optimized angles of rotation we managed to derive exact analytic formulas for the angles of rotation for the case of four wave plates.
A broadband rotator at angle $\alpha $ composed of four half-wave plates is given as
\begin{equation}
\mathbf{J}_{\alpha }(\pi )=J_{\theta _{1}}(\pi )J_{\theta _{2}}(\pi )J_{\theta _{3}}(\pi )J_{\theta _{4}}(\pi ),
\end{equation}
with
\begin{subequations}
\begin{align}
\theta _{1} &=\alpha /8, \\
\theta _{2} &=\pi /2-\alpha /8, \\
\theta _{3} &=3\pi /2-3\alpha /8, \\
\theta _{4} &=\pi -5\alpha /8.
\end{align}
\end{subequations}
Unfortunately, we could not find analytical formulas for longer sequences of wave plates, but we suspect such formula may exists for 8 and 12 wave plate series.

\section{Experiment}

\subsection{Experimental setup}
Experimental investigation of the composite linear polarization rotator described above was performed by analysing polarization of the passed light beam.
The light source was a halogen lamp TUNGSRAM powered by a 6V d.c. power supply.
It covered a broad continuous spectral region from 400 nm to 1100 nm.
A set of two irises and two planoconvex lenses, respectively $L_{1}, f_{1} = 20$ $mm$ and $L_{2}, f_{2} = 150$ $mm$, were used for producing a collimated white light beam with a diameter of about 3 mm, see Fig.~\ref{fig1}.
The light was linearly polarized in the vertical plane by a polariser $P_{1}$, borrowed from a Lambda-950 spectrometer (Perkin Elmer, Glan-Tayler type, spectral range of 210-1100 nm). Thereafter it was directed through the investigated composed rotator.
The polarization of the output beam was analysed by a second polariser $P_{2}$, the same type as the $P_{1}$.
Subsequently, it was directed  by a lens $L_{3}, f_{3} = 15$ $mm$ to the optical fibre $F$ of the spectrometer AvaSpec - 3648 Fiber Optic Spectrometer controlled by the proprietary software AvaSoft - version 7.5.

The linear polarization rotators were built as stacks of an even number of ordinary multi-order quarter-wave plates (WPMQ10M-780, Thorlabs Inc.), the fast axes of which were rotated at the respective theoretically calculated angles given in Tables \ref{table1} and \ref{table2}.
Each wave plate was $1"$ and was assembled onto a RSP1 (Thorlabs Inc.) rotation mount.
This mounting allows rotation at $360^{\circ}$ and accuracy of the rotation angle $1^{\circ}$.
The multi-order wave plates are designed to be 11.25 waves and serve as a quarter wave plates at 780 nm, while at 763 nm they perform like half-wave plates.
Here we focus at the region where the wave plates act as half-wave plates.

\subsection{Measurement procedure}
The measurement started with taking the reference signal before applying the rotation procedure.
The measurement's series of steps were similar to those taken in Ref.~\cite{Dimova03}.
We saved the dark and reference spectra.
The dark spectrum was taken with the light suspended and was further used to automatically correct for hardware offsets.
The reference spectrum was taken with the light source on and using a blank specimen instead the specimen under test.
For this purpose the fast axes of the wave plates were aligned with the axes of the polarisers $P_{1}$ and $P_{2}$.
The integration time of 15 $ms$ and the 1000 data averaging were kept the same during all following measurements.
We looked at the transmittance mode of the spectrometer which for the reference signal is 100\%.
For each investigated set of $N$ wave plates (WPs), where $N$ is $2, 4, 6, 8, 10$ WPs, we took the respective reference signal.

The reference signal presents zero rotation of the linearly polarized light.
The next step was to realize the linear polarization rotator by rotating the fast axis of each WP at the respective angle $\theta_{n}$.
The measurements were accomplished at 6 rotation angles of the polarization $\alpha \in \lbrace15^{0}, 30^{0}, 45^{0}, 60^{0}, 75^{0}, 90^{0}\rbrace$ for each set.
The analysis was done by rotating the polariser $P_{2}$ at the target angle $\alpha$.

\subsection{Experimental results}

\begin{figure}[tbh]
\centerline{\includegraphics[width=1\columnwidth]{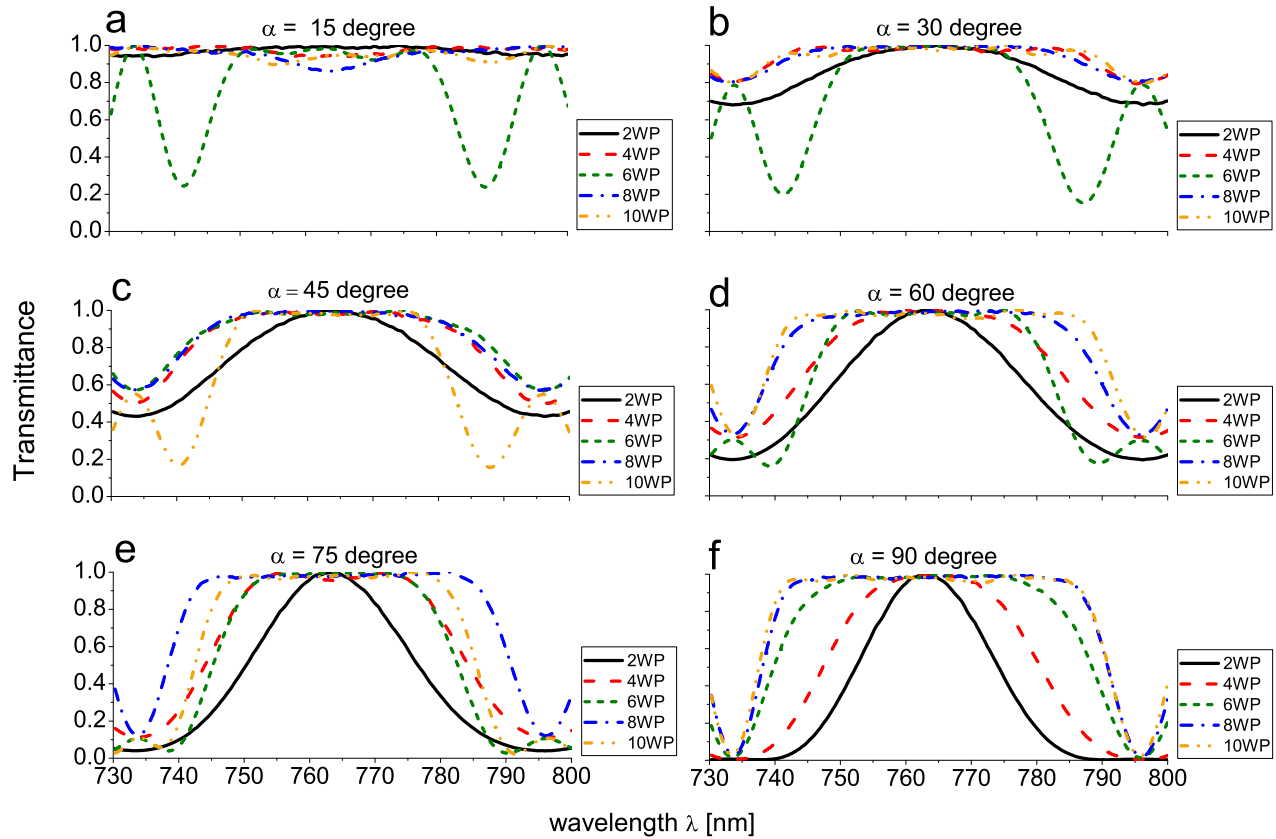}}
 \caption{Measured transmittance vs wavelength for broadband polarization rotator, consisting of different number of half-wave plates  $N$ ($2,4,6,8$ and $10$).
The rotation angle $\alpha$ is denoted in each frame.
}
\label{fig2}
\end{figure}

Angle tunable broadband polarization rotators comprising 2, 4, 6, 8, and 10 wave plates were experimentally demonstrated and the results correspond very accurately to the predicted theoretically ones.
In Fig.~\ref{fig2} we present the experimentally measured spectra and each group of curves demonstrates the rotation effect of the chosen sets of wave plates at each angle $\alpha$,  $\alpha \in \lbrace15^{0}, 30^{0}, 45^{0}, 60^{0}, 75^{0}, 90^{0}\rbrace$.
The broadening effect is proved by comparison with the set of two WPs.
The experimental results show flat maxima in large spectral interval that are maintained for the different rotator angles.
One can see that in principal the broadening of the bandwidth increases with the number on half-wave plates in the set. However the best implementation of the rotator happen when polarization rotators comprise of 4 and 8 wave plates, therefore we have the reason to believe that combination of multiple of 4 wave plates give optimal performance.
It is also evident from Fig. \ref{fig2} that the rotator is more robust against variations in the wavelength in the case of small rotation angles compared to large rotation angles.
This is a novel feature of this composite rotator in contrast to previous composite rotators \cite{Rangelov,Dimova}, where almost the same spectral shape was observed.
In this sense our composite rotator with four half-wave plates outperforms the previous rotator \cite{Rangelov,Dimova} with six half-wave plates up to rotation angles $\pi /4$.\section{Conclusion}
In this paper, we introduced and experimentally demonstrated a novel type of broadband composite polarization rotator that can rotate the polarization plane of a linearly-polarized light at any chosen angle.
The experimental results show strong broadening of the bandwidth of the polarization rotator in case of 4 and 8 half-wave plates and moderate broadband increase in other cases as the number of half-wave plates grow.

\section*{Acknowledgment}
This work was supported by the Bulgarian Science Fund Grant No. DN 18/14 and EU Horizon-2020 ITN project LIMQUET (contract number 765075).


\end{document}